# Erbium Probes of Magnetic Order in a Layered van der Waals Material

Guadalupe García-Arellano[1], Kang Xu[1], Arun Ramanathan[3], Jiayi Li[3], Gabriel I. López-Morales[4], Xavier Roy[3], Cyrus E. Dreyer[4,5], and Carlos A. Meriles[1,2,*]

[1]Department of Physics, CUNY-City College of New York, New York, New York 10031, USA.
[2]CUNY-Graduate Center, New York, New York 10016, USA.
[3]Department of Chemistry, Columbia University, New York, New York 10027, USA.
[4]Department of Physics and Astronomy, Stony Brook University, Stony Brook, New York, 11794-3800, USA.
[5]Center for Computational Quantum Physics, Flatiron Institute, 162 5th Avenue, New York, New York 10010, USA.

**ABSTRACT:** There is growing interest in characterizing magnetic order and dynamics in two-dimensional magnets, yet most efforts to date rely on external probes that interrogate the sample from tens of nanometers away and inevitably average over that length scale. Here we use internal, lattice-embedded $Er^{3+}$ defects in CrSBr as atomic-scale probes, accessing their telecom-band photoluminescence with spectroscopy and temperature-dependent confocal imaging to read out magnetism from within the material. At room temperature we observe narrow, long-lived photoluminescence (PL) lines in the telecom band, characteristic of erbium emitters. Upon cooling to 3 K and reheating, the $Er^{3+}$ PL intensity and excited-state lifetime display pronounced thermal hysteresis with a minimum near 132 K, at the reported antiferromagnetic (AFM) transition of CrSBr. Remarkably, we observe magnetic signatures persisting over a broader temperature range than expected from bulk benchmarks, suggesting nanoscale magnetic order that locally survives beyond the nominal phase boundary. Further, a moderate in-plane field of 0.3 T shifts the PL minimum by +8 K, which we tentatively associate to field-biased ferromagnetic correlations.

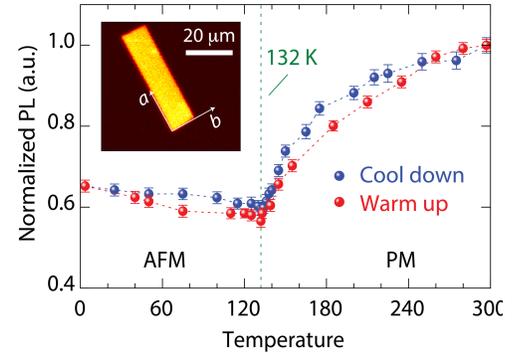

Crystalline van der Waals (vdW) magnets provide a fertile platform to study correlated excitations and engineer hybrid magnonic–optoelectronic devices at the two-dimensional (2D) limit. Among them, the magnetic semiconductor chromium sulfide bromide (CrSBr) is distinctive, as it combines a direct electronic gap in the near-infrared and a comparatively high magnetic ordering temperature[1]; further, it exhibits quasi-one-dimensional (quasi-1D) electronic character that drives highly anisotropic transport and optics[2]. In bulk and few-layer form, CrSBr shows A-type antiferromagnetism (AFM) with a Néel temperature $T_N \approx 132$ K and large in-plane anisotropy, as well as good exfoliability and air stability, making it an attractive host for exploring mixed 1D/2D physics and magneto-optical coupling[3,4].

A growing body of work shows that CrSBr's excitonic response is tightly correlated with its magnetic state[5-7]. Quasi-1D Cr–S chains along the $b$-axis underpin extreme transport anisotropy and polarization-dependent PL, with ultraclean, narrow excitons and pronounced exciton–phonon/polariton effects[2,3]. In this mixed-dimensional setting, excitonic resonances track field- and temperature-driven reorganizations of magnetic order[8], including a 30–40 K 'hidden-order' window[3,9] as well as direct exciton–magnon coupling and tunable hybridization in CrSBr[10,11]; these magneto-optical contrasts furnish practical readouts of magnetic state in few-layer CrSBr[5,12,13].

Adding to various characterization methods[4,5,7,14], prior work leveraged individual nitrogen-vacancy (NV) centers in diamond as atomic probes to image CrSBr magnetism with high spatial resolution[15]. Complementary ensemble NV $T_2$ noise magnetometry has quantified critical slowing-down and extracted critical exponents[16]. These approaches deliver quantitative nanoscale maps of stray fields, magnetic anisotropy, and phase coexistence in monolayer and bilayer samples; however, because the NV is external to the sample, the spatial averaging length is typically set by an NV–sample standoff on the order of tens of nanometers, which inevitably averages magnetic textures over that scale.

As demonstrated recently for Yb:$CrI_3$[17], an alternative approach is to embed the atomic sensor within the magnetic material, an idea we adapt here to CrSBr by resorting to optically active erbium ($Er^{3+}$) defects. $Er^{3+}$ centers provide telecom-band optical transitions with magnetic-dipole character and nanoscale (atomic-site) locality, enabling sensitivity to the immediate magnetic environment without the geometric averaging intrinsic to external probes. Using ion implantation and cryogenic optical microscopy, we observe narrow telecom emission lines. Upon cooling and warming, the $Er^{3+}$ PL intensity displays a pronounced thermal dependence with a minimum at 132 K, the bulk $T_N$. Signatures of residual local magnetization persist over a broader temperature window than expected from bulk benchmarks, suggesting nanoscale magnetic order that survives beyond the nominal phase boundary. This notion is further supported by the observation of thermal hysteresis in the PL response, manifesting as reproducible intensity differences correlated with the sample cooling history. Moreover, we find that a moderate in-plane field of 0.3 T shifts the apparent transition approximately by +8 K. Beyond demonstrating the use of Er ions as local probes of magnetic order, these results portend a route for interfacing spin-polarized Er excitations with CrSBr excitons and magnons, an integrated platform for spin–photon–magnon transduction in a 2D semiconductor[2,10,18].

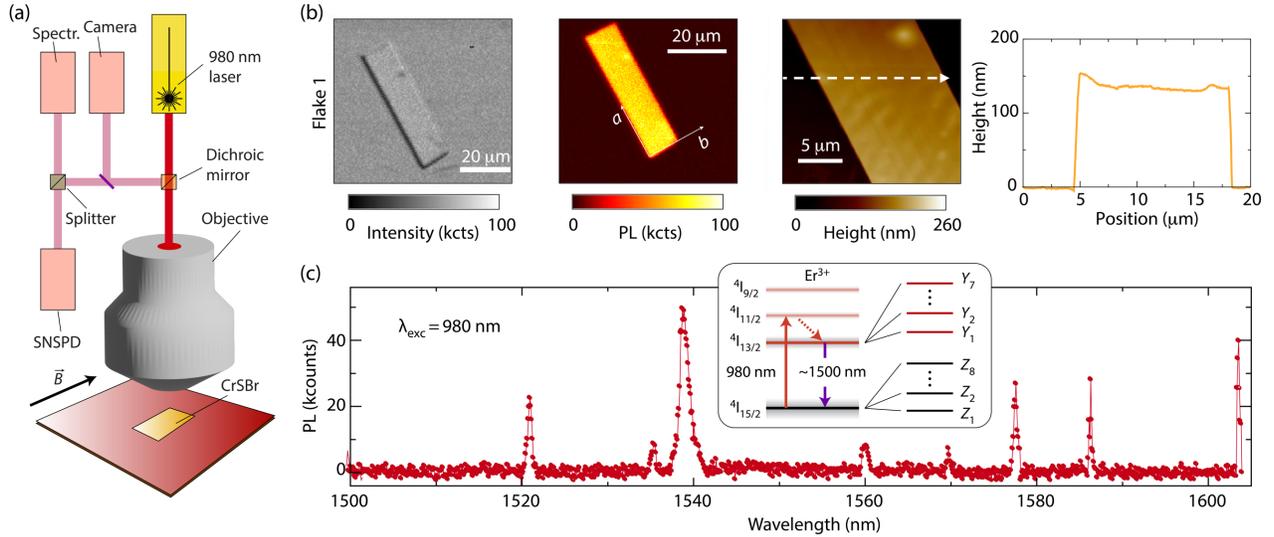

**Figure 1: Er implantation and room-temperature telecom emission from Er:CrSBr.** (a) Schematic of the experimental setup; the magnetic field $\vec{B}$ is produced by a permanent magnet (not shown). A cryo-workstation allows us to vary temperature from ambient down to 3.5 K. (b) Overview of an Er-implanted CrSBr flake (here referred to as Flake 1): Reflected-light micrograph (left), confocal PL map integrated in the telecom band (center-left), atomic force microscopy topography (center-right), and height profile along the dashed line (right); $a$ and $b$ denote the CrSBr crystallographic axes. (c) Representative room-temperature PL spectrum showing multiple narrow $Er^{3+}$ lines in the telecom band with prominent features near 1520, 1540, and 1560 nm with additional lines extending toward ~1.60 μm. All experiments in (a) and (b) are performed at room temperature. SNSPD: Superconducting nanowire single-photon detector. Spectr.: Spectrometer.

## RESULTS AND DISCUSSION

In our experiments we use a home-built confocal microscope with below-gap laser excitation at 980 nm. The pump is focused onto the sample through a high-NA objective; emitted photons in the telecom band are separated by a dichroic/long-pass stack and directed either to a grating spectrometer with an InGaAs camera or to a superconducting nanowire single-photon detector (SNSPD). A cryogenic workstation provides temperature control from ambient down to ~3.5 K, and a permanent magnet next to the sample supplies the static field $\vec{B}$ used in later measurements (Fig. 1a, see also Supplementary Information (SI), Sections I and II).

The samples we investigate were produced by exfoliation from a high-purity CrSBr crystal followed by ion implantation (10 keV, $10^{13}$ ions $cm^{-2}$). In the absence of preceding work on Er emission in this host, we first characterize representative flakes at room temperature. To facilitate detection—especially demanding in time-resolved and spectroscopy measurements, see below—we focus on relatively thick flakes ($\geq$ 80 nm) where the PL brightness is highest. Figure 1b compiles a reflected-light image, a confocal PL map integrated over the 1.5–1.6 μm band, and the corresponding atomic force microscopy topography with an accompanying line profile. The flake we present here features a characteristic rectangular geometry, elongated along the $a$-axis. The PL is co-localized with the crystal and follows the flake geometry. In this particular illustration, the emitted PL is homogeneous throughout the crystal, consistent with the uniform thickness of the flake.

High-resolution spectroscopy (Fig. 1c) reveals multiple resonances throughout the telecom region. The spectral structure is consistent with transitions connecting the $Er^{3+}$ $^4I_{13/2}$ and $^4I_{15/2}$ manifolds in a crystalline host[19]. Compared with prior studies of telecom emission in Er:WS$_2$[20], these room-temperature PL lines are broader indicating greater lattice heterogeneity, potentially from ionic size mismatch and/or multiple microscopic configurations of the crystalline and/or magnetic environments (see below). Further, besides the hallmark peaks at 1520 and 1540 nm, the spectrum also contains multiple longer-wavelength peaks, indicating crystal-symmetry-enabled relaxation and emission channels not seen in WS$_2$. The distinct behavior of Er ions in this host extends to the excitation and emission dipoles, colinear in WS$_2$ but rotated by 60 degrees in CrSBr (SI, Section III).

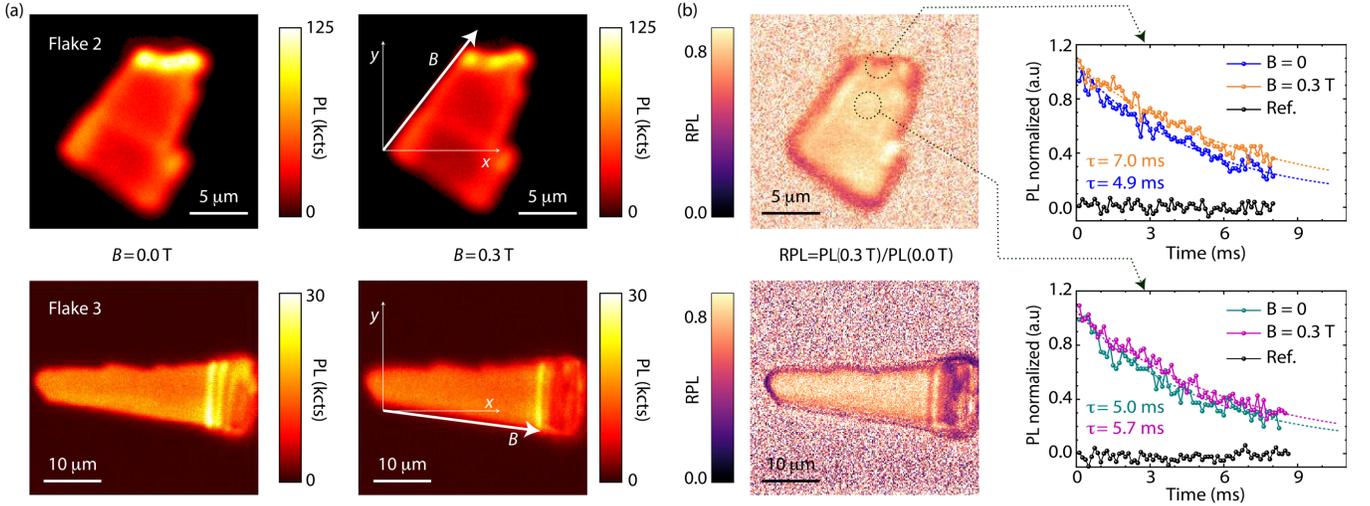

**Figure 2: Magnetic-field–induced quenching of Er³⁺ photoluminescence in Er:CrSBr.** (a) Telecom-band PL maps from two implanted Flakes (2, 3) at $B = 0$ and with an in-plane field $B = 0.3$ T parallel to the $a$-axis. In both flakes the field reduces the overall PL intensity. (b) Ratio images RPL = PL(0.3T)/PL(0T) highlighting spatially nonuniform quenching. In Flake 2 the suppression is strongest near the perimeter; in Flake 3 it extends to selected bright regions. The edge- or region-selective response is not universal but recurs across samples. (Right-hand side inserts) PL lifetime with and without magnetic field as measured near the top edge of Flake 2 (top) and within its bulk area (bottom); dashed lines indicate exponential fits. Ref: Reference signal in the absence of optical excitation. All data at room temperature and PL integrates emission above 1500 nm.

With the basic spectroscopy established, we next examined how an in-plane magnetic field influences the Er signal at room temperature. Figure 2 compares telecom-band PL maps of two implanted flakes with $B = 0$ and $B = 0.3$ T approximately colinear with the magnetic intermediate $a$-axis. In both cases, we find that the presence of magnetic field reduces the PL intensity by about 20%—both under ambient and cryogenic conditions, see below—a dependence that makes Er a practical reporter of magnetic response.

The microscopic origin is not yet clear, to a large extent because we presently don't know the microscopic crystallographic structure around Er emitters. One plausible mechanism, however, could involve a reduction of the dipole oscillator strength due to Zeeman mixing, a route potentially facilitated by the strong magnetic dipole component of erbium's telecom emission lines[21,22]. In this scenario, the magnetic field at the Er site perturbs crystal-field levels in a way that reduces the magnetic-dipole radiative rate; assuming the magnetic field leaves the non-radiative decay channels unaffected, the latter leads to longer emitter lifetimes and reduced quantum efficiency —i.e., less PL—as experimentally observed. Notably, we find the PL change is weaker for fields aligned along the $b$-axis and virtually null pointing out-of-plane (SI, Section III), i.e., the Er readout is selectively sensitive to in-plane field components, an important consideration when interpreting the impact of CrSBr's intrinsic magnetization (see below).

We caution that PL maps such as those presented in Fig. 2a may also be impacted by photonic effects: Specifically, an in-plane field can modify the host's magneto-optical response and, with it, the local density of optical states (LDOS), as slab-guided/leaky modes are cut off or redirected. Note that if $B$ changes the birefringence/dichroism of CrSBr, the edge region—where mode structure is already most sensitive—will show the largest field-induced PL change. This view is generally consistent with the ratio maps in Fig. 2b, where quenching often concentrates at edges or along high-PL ridges (although the effect is not universal among flakes, vanishing in ≲ 50 nm thick flakes).

As shown in the right-hand side insets of Fig. 2b, PL dimming is accompanied by longer excited state lifetimes, an anticorrelation we find invariably throughout our experiments. Assuming the non-radiative channels remain unchanged, this combined dependence points to a reduction of the Er radiative rate in the presence of an external field. Note that, similar to the PL intensity, the lifetime change can also be amplified by photonic effects, as seen when comparing the bulk and near-edge responses.

Pending a full mechanistic assignment, the established PL/lifetime–field coupling provides a sensitive reporter that can be leveraged to map the evolution of magnetism in CrSBr across the Néel transition. Indeed, cooling the implanted flakes from room temperature reveals a pronounced suppression of Er³⁺ telecom PL that deepens into a clear minimum around $T_N \approx 132$ K (Figs. 3a through 3c); at this Néel point, the system transitions from paramagnetic to an A-type antiferromagnetic phase, featuring ferromagnetically aligned spins within each vdW layers that are antiferromagnetically coupled between adjacent layers[4,5]. Accompanying this change in PL intensity, we also measure a mirrored temperature dependence of the emission lifetime, exhibiting a clear maximum at $T_N$ (Figs. 3d and 3e).

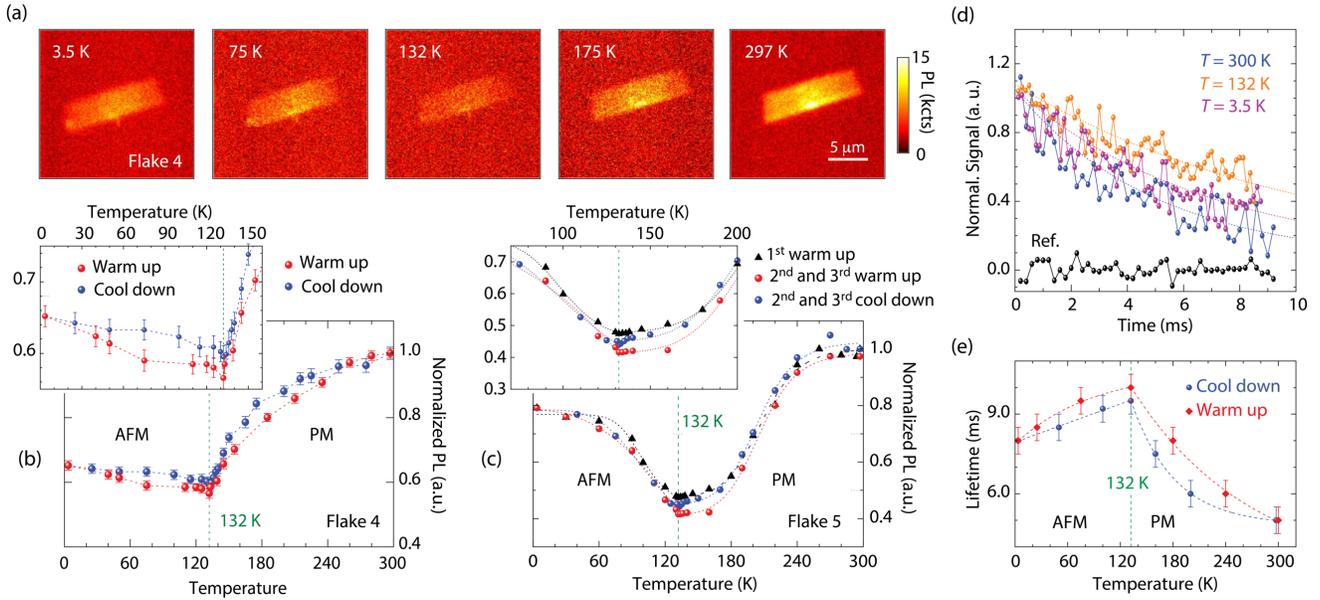

**Figure 3. Er³⁺ PL and lifetime across the Néel transition.** (a) Telecom-band PL images for Flake 4 (~70 nm thick) acquired at selected temperatures during a temperature cycle (cooling and reheating, as indicated). The emission weakens markedly near the Néel temperature $T_N \approx 132$ K. (b) Flake-integrated PL versus temperature for cooling (blue) and warming (red), revealing pronounced thermal hysteresis with a minimum centered near $T_N$. Inset: magnified view of the 90–160 K range. (c) Normalized PL from repeated cycles on a thicker flake (Flake 5, ~600 nm), highlighting the convergence and ultimate reproducibility of the dip near $T_N$ and the hysteretic envelope. (d) Representative time-resolved PL traces at several temperatures (symbols) for Flake 4 with single-exponential fits (dashed). The excited-state lifetime increases on approaching $T_N$. A reference trace from an off-Er region is shown in black. (e) The extracted PL lifetime peaks near $T_N$ and decreases away from the transition (arrow indicates sweep direction). No external magnetic field is present throughout these measurements.

Consistent with the observations in Fig. 2, here we take the Er signal to be an inverse reporter of the net local, in-plane magnetic field at the defect site: When the net field grows, the effective radiative rate drops and the PL intensity decreases. Importantly, each Er probes its immediate lattice environment so the signal reflects nm-scale magnetic imbalances rather than mesoscopic averages. Further, with millisecond excited-state lifetimes, the Er center is sensitive to quasi-static fields on that timescale—as opposed to fast fluctuation—so we interpret the readout to primarily reflect how well the two AFM sublattices cancel at the Er location and how much uncompensated magnetic moment resides in its immediate nm-scale neighborhood.

Viewed through this lens, our results suggest a rich temperature dependence of the local magnetic order. At $T_N$, interlayer phase locking is weakest and the crystal hosts short-correlated AFM fragments and mobile domain walls[4,7,9,15,23]; correspondingly, incomplete sublattice cancellation and wall/edge/defect moments yield the largest net local field, producing the pronounced PL minimum (with lifetime maximum). Below $T_N$, long-range AFM order gradually strengthens and sublattice cancellation at the Er site improves; domain walls freeze and the local field decreases, so the emission rate and PL increase. This recovery is more pronounced in the thicker flake (~600-nm-thick Flake 5), consistent with more bulk-like order (we have observed similar thickness-dependent PL trends when comparing the response from different areas in non-uniform flakes, see SI, Section III).

Using the PL dependence in Fig. 2 as a rough proxy, the fractional dimming at $T_N$ corresponds to an effective in-plane field $B_{eff}$ of a few hundred mT at the Er site; we emphasize this is a phenomenological equivalence, not a direct measurement of the microscopic dipolar field. Remarkably, reduced PL intensity persists over a broad temperature range *above* $T_N$, indicating residual local magnetic fields or short-range order sensed at the defect site; the high temperature shoulder extends further in the thicker flake, consistent with longer correlation lengths, and greater metastability[3-5,24,25].

It is worth contrasting this response with prior NV-based studies—both single-spin scanning magnetometry[15] and ensemble $T_2$ noise spectroscopy[16]—where the sensed stray fields are orders of magnitude weaker and critical behavior comparatively concentrates near the transition temperature. We attribute these differences to the external nature of these sensors, necessarily averaging over nanometric stand-off distances. For example, in scanning NV experiments on CrSBr, the effective NV–sample spacing that sets spatial resolution is greater than ~10 nm, which inherently smooths magnetic textures on that scale; Er, by contrast, is an internal, atomic-site probe that reports on fields within a few ångströms of its lattice position, a sensing locality akin to muon spin relaxation (μSR) spectroscopy[9].

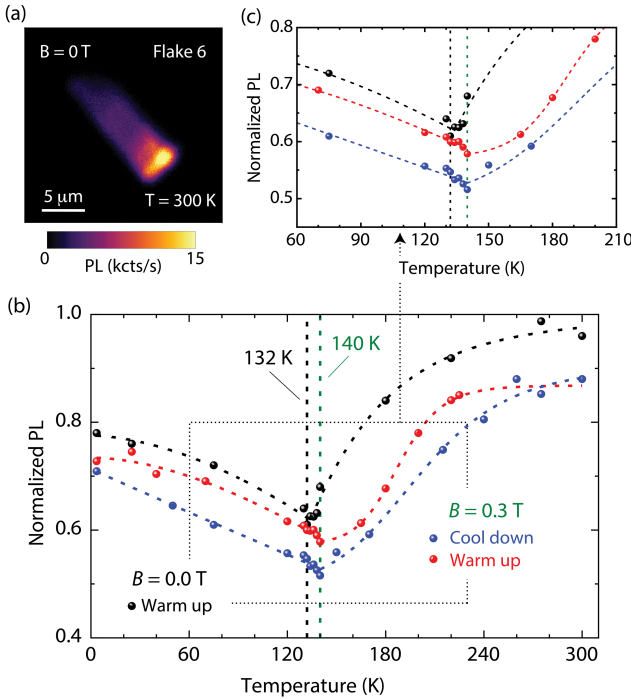

**Figure 4. Dependence of $T_N$ on applied magnetic field.** (a) Telecom-band PL map of Flake 6 (~100 nm thick) at room temperature and zero field. (b) Integrated PL versus temperature at $B = 0.0$ T (black, warm-up) and with an in-plane field $B = 0.3$ T (blue, cool-down; red, warm-up). The field uniformly reduces the PL amplitude and shifts the PL minimum — used here as a proxy for the transition — from ~132 K (zero field) to ~140 K. Dashed curves are guides to the eye. (c) Zoom of the boxed region in (b), highlighting the field-induced dimming and the upward shift of the minimum. In (b) and (c), the PL intensity has been normalized relative to its value at 300 K and 0.0 T.

A separate hallmark in the results of Fig. 3 is the thermal hysteresis observed on cycling the temperature, both in the PL intensity and lifetime. When warming from low $T$ (red dots in Fig. 3b), the PL remains comparatively dimmer than on cooling at the same temperature (blue dots), suggesting persisting magnetic order upon warming—i.e., a larger remnant order parameter carried upward from the AFM phase—before relaxing toward the paramagnetic state. A further nuance is that the hysteresis trains before it settles. In the thicker flake of Fig. 3c, for example, the first warm-up (black triangles) defines a baseline; the subsequent cool-down (second cool-down, blue circles) then lies below that first warm-up—the opposite trend from Fig. 3b, where cool-down curves sit above warm-up curves at the same $T$. Only after additional cycles ($2^{nd}$ warm-up → $3^{rd}$ cool-down → $3^{rd}$ warm-up) do the traces converge to a repeatable loop with the standard ordering (cool-down above warm-up) observed in Fig. 3b. This history dependence points to slow domain/wall reconfiguration and pinning–depinning that is especially pronounced in thicker flakes; repeated thermal excursions anneal the metastable landscape, after which the system reproducibly follows the same microscopic pathway through the AFM–PM transition.

With the zero-field behavior stabilized after a few training cycles, we are positioned to ask how an in-plane external field perturbs the observed response. We address this question in Fig. 4 where we map the temperature dependence of the $Er^{3+}$ PL in the case where a moderate $B$ field aligns in plane; in this example the field is approximately parallel to the crystalline $a$-axis. Adding to an overall PL dimming—consistent with the results in Fig. 2—we unexpectedly find that the field also alters the transition signature: The PL minimum shifts from 132 K at $B = 0$ to ~140 K at $B = 0.3$ T. The shift we observe is surprising in an antiferromagnet: naively, one expects an applied field to depress, not raise, the AFM transition scale. An illustration is the intermetallic antiferromagnet $Yb_3Pt_4$, where $T_N$ decreases roughly monotonically with field reflecting a Zeeman-driven collapse of the AFM state[26].

One possible reading of our data—better matched to what we know about CrSBr[4,14]—is therefore not that $T_N$ increases, but that an in-plane field biases critical fluctuations to promote ferromagnetic–like correlations. In this picture, an applied $B$ along the plane partially aligns these fluctuations, seeding finite ferromagnetic (FM) puddles that are local rather than long-range. As an internal, nm-scale sensor of the net local field, Er registers this field-polarized pocket as a deeper PL suppression that extends to higher apparent temperature. Future work combining μSR and neutron scattering could shed light on these dynamics.

## CONCLUSION

In summary, we have shown that erbium can act from within CrSBr as an informative, telecom-band optical probe of magnetism. Room-temperature spectra reveal narrow $Er^{3+}$ lines, whose amplitude responds sensitively to magnetic perturbations: A modest in-plane field dims the PL, while temperature sweeps uncover a pronounced minimum near $T_N$ together with a broad high-temperature shoulder that signals residual local order above the bulk transition. Cycling the temperature produces clear hysteresis—evidence that the system "remembers" low-$T$ order on reheating—and, with the field applied, the PL minimum shifts upward by ~8 K. Combined, these behaviors are consistent with an internal, site-specific sensor, revealing magnetic textures and dynamics that external probes can average away.

Interestingly, the Er PL remains comparatively flat across 30–40 K, with no distinct anomaly. This behavior suggests that the intrinsic low-temperature crossover reported in CrSBr[9]—often described as a hidden-order/dynamic magnetic regime—does not substantially modify the net local field at the Er site, implying that near-site sublattice cancellation is largely preserved through this range (i.e., the local order sensed by Er remains effectively unchanged).

Although we have treated Er as a largely passive probe with negligible back-action (implantation doses here are moderate), systematic studies versus Er concentration will be important to assess whether the dopants themselves perturb CrSBr's magnetic dynamics, e.g., by modifying anisotropy, and pinning/domain kinetics. Along related lines, further site-resolved spectroscopy and first-principles calculations are needed to pin down the Er lattice position(s) and symmetry, and to quantify how local magnetic fields modify oscillator strengths and branching ratios. Because Er also offers a telecom-compatible spin qubit, integrating coherent control (ODMR, optical pumping) with waveguides or microcavities could yield on-chip, fiber-ready readout of layered magnets, and enable spin–photon–magnon interfaces[18] linking magnetic excitations in CrSBr to long-distance optical networks.

## AUTHOR INFORMATION

**Corresponding author:**
[†]E-mail: cmeriles@ccny.cuny.edu

## DATA AVAILABILITY

The data that support the findings of this study are available from the corresponding author upon reasonable request.

## SUPPORTING INFORMATION

Contains details on the experimental setup, measurement protocols and conditions, as well as optical spectroscopy and lifetime measurements under cryogenic conditions.

## ACKNOWLEDGEMENTS

The authors thank Vinod Menon for helpful discussions. G.G.A. and C.A.M. acknowledge support from the Department of War, Award W911NF-25-1-0134. C.A.M. also acknowledges support from the U.S. Department of Energy, Office of Science, National Quantum Information Science Research Centers, Co-design Center for Quantum Advantage (C2QA) under contract number DE-SC0012704. K.X acknowledges support from the National Science Foundation via grant NSF-2328993. CED and GILM acknowledge support from the National Science Foundation under award DMR-2237674. Synthesis, structural and magnetic characterization efforts at Columbia were supported as part of the Programmable Quantum Materials, an Energy Frontier Research Center, funded by the US Department of Energy (DOE), Office of Science, Basic Energy Sciences, under award no. DE-SC0019443. All authors also acknowledge access to the facilities and research infrastructure of the NSF CREST IDEALS, grant number NSF-2112550. We also acknowledge the use of facilities and instrumentation supported by the NSF through the Columbia University, Materials Research Science and Engineering Center (Grant No. DMR-2011738). The Flatiron Institute is a division of the Simons Foundation.

# Erbium Probes of Magnetic Order in a Layered van der Waals Material

Guadalupe García-Arellano[1], Kang Xu[1], Arun Ramanathan[3], Jiayi Li[3], Gabriel I. López-Morales[4], Xavier Roy[3], Cyrus E. Dreyer[4,5], and Carlos A. Meriles[1,2,*]

[1]*Department of Physics, CUNY-City College of New York, New York, New York 10031, USA.*
[2]*CUNY-Graduate Center, New York, New York 10016, USA.*
[3]*Department of Chemistry, Columbia University, New York, New York 10027, USA.*
[4]*Department of Physics and Astronomy, Stony Brook University, Stony Brook, New York, 11794-3800, USA.*
[5]*Center for Computational Quantum Physics, Flatiron Institute, 162 5th Avenue, New York, New York 10010, USA.*

**Content**

# I. Sample characterization

## I.a Synthesis of bulk CrSBr crystals

The synthesis follows a previously established procedure[1]. The following reagents were used as received unless otherwise stated: chromium powder (99.94%, −200 mesh, Alfa Aesar), sulfur pieces (99.9995%, Alfa Aesar), bromine (99.99%, Aldrich) and chromium dichloride (anhydrous, 99.9%, Strem Chemicals). To begin, $CrBr_3$ was synthesized from Cr (1.78 g, 34.2 mmol) and $Br_2$ (8.41 g, 52.6 mmol) with one end of the tube held at 1,000 °C and the other end at 50 °C with a water bath. Of note, one end of the tube must be maintained below 120 °C to prevent the tube from exploding from bromine overpressure. Chromium (0.174 g, 3.35 mmol), sulfur (0.196 g, 6.11 mmol) and CrBr3 (0.803 g, 2.75 mmol) were loaded into a 12.7-mm-outer-diameter, 10.5-mm-inner-diameter fused silica tube. The tube was evacuated to ~30 mTorr and flame sealed to a length of 20 cm. It was then placed into a computer-controlled, two-zone, tube furnace. The source side was heated to 850 °C over 24 h, allowed to soak for 24 h, heated to 950 °C over 12 h, allowed to soak for 48 h and then cooled to ambient temperature over 6 h. The sink side was heated to 950 °C over 24 h, allowed to soak for 24 h, heated to 850 °C over 12 h, allowed to soak for 48 h and then cooled to ambient temperature over 6 h. The crystals were cleaned by soaking in a 1 mg ml$^{-1}$ $CrCl_2$ aqueous solution for 1 h at ambient temperature. After soaking, the solution was decanted and the crystals were thoroughly rinsed with deionized water and acetone. Residual sulfur residue was removed by washing with warm toluene.

## I.b Sample fabrication and implantation details

Flakes were prepared by mechanically exfoliating high-purity CrSBr onto 285-nm $SiO_2$/Si (HF-cleaned just prior to transfer) and screened in bright-field reflection. All flakes were implanted simultaneously in a single broad-beam step (Cutting Edge Ions) at low current density, using 10 keV at 7 deg from the normal to the surface to avoid channeling, with a nominal fluence of ~1×10$^{13}$ cm$^{-2}$. After implantation, samples were optionally annealed (see below), and mounted in the cryo-workstation with a low-fluorescence, vacuum-compatible resist; room-temperature PL imaging confirmed Er activation, and regions with obvious dose nonuniformity were excluded from quantitative analysis.

To estimate the dopant depth, we first used SRIM (Stopping and Range of Ions in Matter), the community's go-to package for ion-range calculations; for CrSBr ($\rho \approx 4.3$–$4.5$ g cm$^{-3}$) SRIM predicts a shallow projected range, with average depth of only ~8.5 nm and straggle $\lesssim$ 20 nm (Supplementary Fig. 1a). However, because SRIM treats the target as static, it is ill-suited for heavy ions in heavy, layered substrates, where sputtering, cascade mixing, densification, and stoichiometry evolution during implantation can be significant. In particular, for thin targets and mid-to-heavy ions (approximately $29 < Z < 83$), experiments show that SRIM often underestimates penetration depth, largely due to inaccuracies in electronic stopping and the neglect of channeling[2-5]. The factor of underestimation can range from 2 (as reported for erbium-implanted silicon films[6]) to 10 (as reported in Xe, Ar, N, and O ions implanted in tungsten[7]).

Alternative binary-collision approximation Monte Carlo transport codes—such as IRADINA[8] and TRI3DYN[9]—have been developed to more accurately predict penetration depths. Supplementary Fig. 1b shows the results for TRIDYN, a variant of TRI3DYN that also includes effects like sputtering and evolving target composition, but is better adapted to thicker flakes and is computationally more efficient. We find that TRIDYN shifts the Er distribution markedly deeper (mean $\approx 10^2$ nm) and broadens it (Supplementary Fig. 1b). An experimental cross-check — room-temperature telecom PL versus thickness — rises and then saturates across flakes, showing characteristic thickness of about ~100 nm (Supplementary Fig. 1c). The latter is difficult to reconcile with the shallow SRIM profile, so we adopt the TRIDYN-like distribution as our working estimate.

Finally, we note that 10 keV was the minimum energy available from the implanter, yet this energy is already too large for few-layer CrSBr, for which ions would largely traverse the crystal; accordingly, our study focuses on thicker flakes ($\gtrsim$ 40 nm), where implanted Er remains within the host.

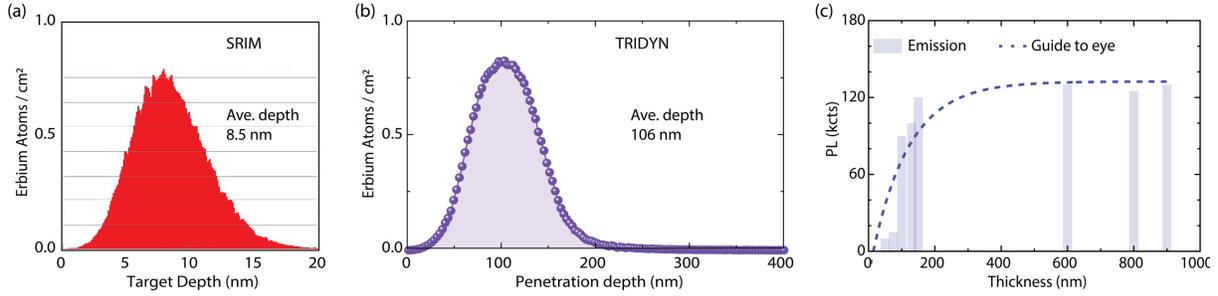

**Supplementary Figure 1: Thickness dependence and implantation depth profiles.** (a) SRIM simulation of Er penetration for implantation conditions: The projected depth distribution peaks within the top tens of nanometers (average depth ≈ 8.5 nm). (b) TRIDYN simulation for the conditions in (a), yielding a deeper projected range with average penetration depth ≈ 106 nm. (c) Room-temperature telecom PL (integrated 1.50–1.60 µm) versus thickness for representative flakes; bars show the mean PL from each flake, and the dashed line is a guide to the eye. The rise and eventual saturation of brightness with increasing flake thickness is qualitatively consistent with the simulations in (b).

## *I.c The effect of mild thermal annealing*

The flakes analyzed in the main text were *not annealed*: Initial telecom PL was already strong, optimal conditions for CrSBr were uncertain, and high-temperature holds risk perturbing magnetic properties. Nonetheless, to benchmark possible benefits/costs, we performed separate anneals on a subset of flakes (argon atmosphere, 1-h hold, Supplementary Fig. 2a). We find a modest PL increase for mild end-temperatures (180 °C), whereas 400 °C (or higher) generally reduces the PL (Supplementary Fig. 2b).

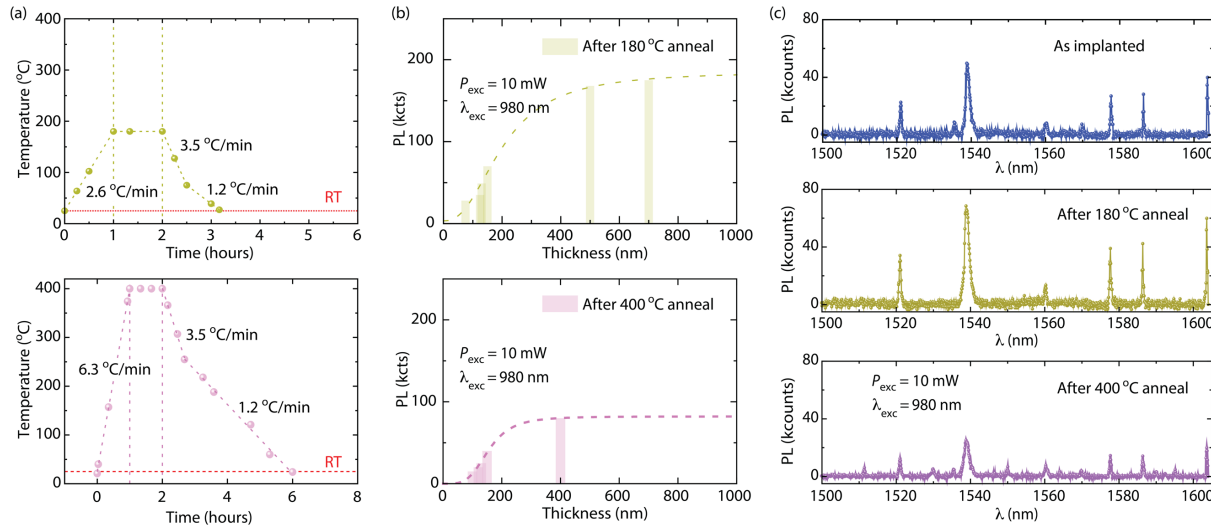

**Supplementary Figure 2: Annealing protocols, thickness dependence, and spectral changes of Er:CrSBr.** (a) Temperature profiles in argon. Top: 180 °C process with ramp ≈ 2.6 °C min$^{-1}$, 1 h dwell, and cooldown ≈ 1.2 °C min$^{-1}$. Bottom: 400 °C process with ramp ≈ 6.3 °C min$^{-1}$, 1 h dwell, and cooldown ≈ 1.2 °C min$^{-1}$. (b) Room-temperature telecom PL versus flake thickness (980-nm, 10 mW excitation) measured after each anneal. A mild anneal at 180 °C increases the PL while preserving the thickness trend (dashed guide to the eye), whereas a 400 °C anneal markedly suppresses the signal across devices. (c) Representative spectra: as-implanted (top), after 180 °C (middle; brighter lines with similar linewidths), and after 400 °C (bottom; overall quenching and appearance of additional weak lines). Gentle annealing seems to repair implantation damage, while high-temperature processing modifies local environments/defect complexes and degrades emission.

Spectroscopically, annealing does not narrow the Er emission lines, indicating limited improvement in local heterogeneity; furthermore, ~400 °C anneals introduce additional weak lines in the telecom spectrum.

These observations align with two considerations. First, low-temperature activation of implanted Er has precedent in our Er:WS$_2$ work, where a gentle anneal "turns on" telecom PL without dramatic linewidth changes (i.e., repairs implantation damage and reconfigures nearby vacancies without blanket reordering[10]). On the other hand, CrSBr's layered, halide-containing lattice is susceptible to stoichiometry drift and defect complex formation at elevated temperature; high-$T$ processing can generate secondary phases and alter local crystal fields[11,12], both of which would diminish brightness and spawn new spectral features rather than sharpen existing ones.

A conservative interpretation of the extra peaks after ~400 °C is the emergence of new Er environments, e.g., Er coupled to halogen/chalcogen vacancies or other implantation-induced complexes, charge-state rebalancing at the site, or incipient diffusion/aggregation inside the flake. Disambiguating these scenarios will require systematic anneal series (time/temperature/ambient), compositional probes (e.g., EDS/EELS, ToF-SIMS), and structure-sensitive spectroscopy (site-selective PL, polarization-resolved PLE). For the purposes of this study — focused on magneto-PL sensing — we therefore avoid post-implant anneals and use non-annealed flakes for the main results.

### *I.d Correlation between photoluminescence and flake topography*

Atomic force microscopy (AFM) scans were acquired in tapping mode under ambient conditions (Bruker Fastscan) using Si tips (TESPA-150 BRUKER) with nominal radius < 8 nm. Height and phase channels were recorded simultaneously. Flake thicknesses were obtained from step-height profiles at naturally cleaved edges and/or shallow scratches made outside optical regions of interest. Representative AFM maps and line-cuts are compiled in Supplementary Fig. 3 along with the corresponding optical and PL images (integration window 1.50–1.60 µm, 980 nm excitation) of all flakes presented in the main text.

Within flakes, local PL variations tend to correlate with topographic features: We observed enhanced PL near sharp, protruding edges and inner ridges/wrinkles, possibly because of better out-coupling, as discussed in the main text. Conversely, the PL is more homogeneous in uniform interiors.

## II Experimental setup

### *II.a Room-temperature optics*

A home-built confocal microscope is driven by a 980-nm diode laser (single-mode fiber–coupled; linewidth < 0.1 nm). The beam is cleaned by a short-pass/neutral-density stack, expanded to fill the back aperture of a high-NA objective (50×, NA = 0.8 air). Typical powers at the sample are 4 mW unless otherwise noted; operation is in the linear (unsaturated) regime. Back-scattered/emitted light is routed through a Thorlabs DMLP-1138 long-pass mirror, two 1500-nm and one 1250-nm long-pass filters to reject the pump. Detection is either (i) an InGaAs camera behind a 300 mm grating spectrometer (900 grooves/mm blazed at 1.6 µm), or (ii) a fiber-coupled ID Quantique superconducting nanowire single photon detector (SNSPD, system jitter ~60 ps; dark counts < 50 s$^{-1}$). A rotatable half-wave plate (HWP) sets the pump polarization; a rotatable analyzer (HWP+polarizer) selects the detection polarization. The sample sits on a motorized *xyz* stage for coarse positioning.

### *II.b Low-temperature operation*

A closed-cycle cryo-workstation (Montana Instruments, optical access, vacuum < 10$^{-5}$ Torr) provides temperature control from 300 K to ~3.5 K (stability ±0.1–0.2 K). The sample is mounted on a low-fluorescence copper chuck with silver paint. Temperature is measured at the sample stage (silicon diode) and cross-checked at the cold finger; reported values refer to the stage sensor. A permanent magnet mounted on a kinematic arm provides a static in-plane field (up to ~0.3 T at the sample). Field magnitude/direction are calibrated once per session by Hall probe; out-of-plane alignment is verified to within ±3–5°. Unless stated, fields are in-plane.

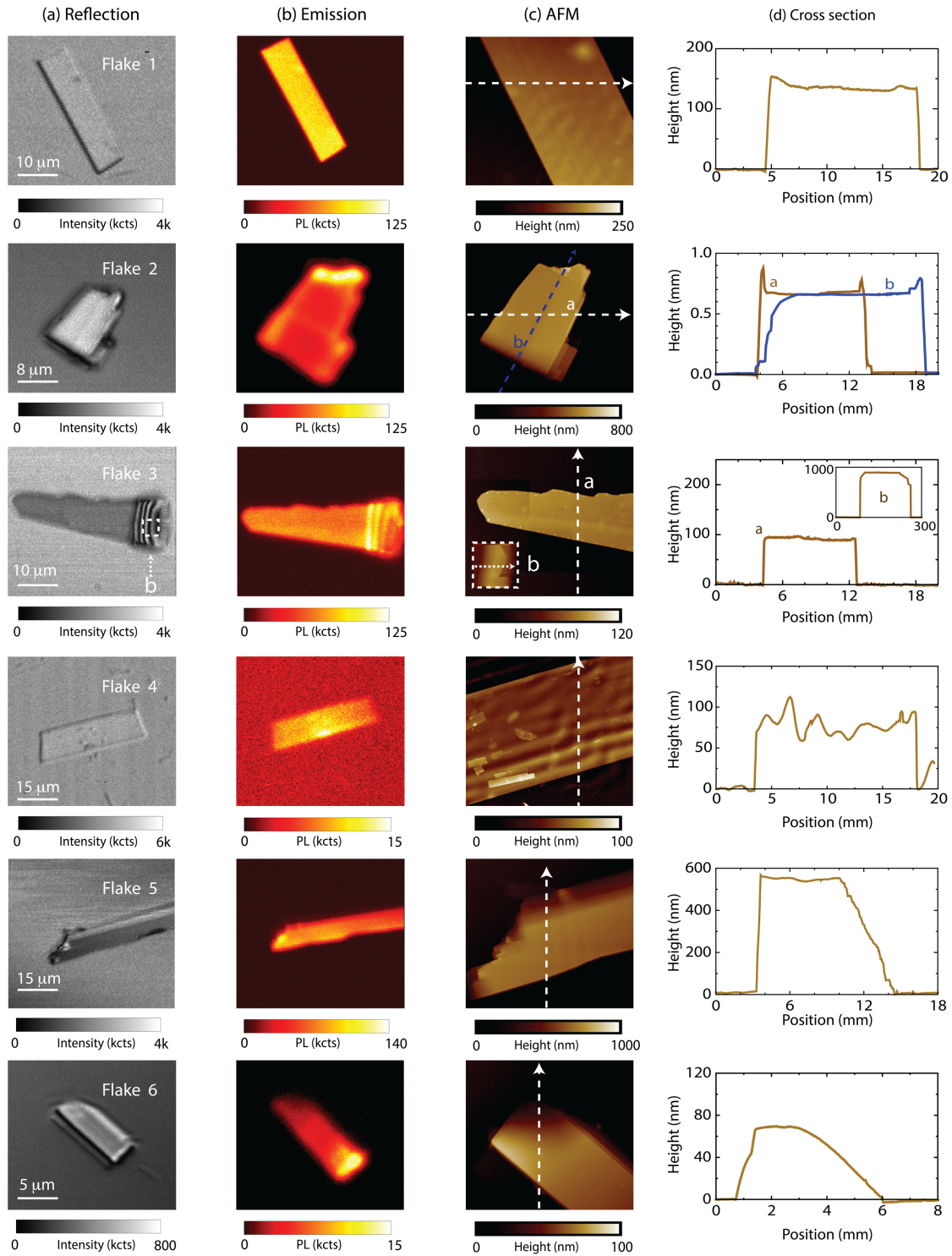

**Supplementary Figure 3: Room-temperature characterization of Er-implanted flakes in the main text.** For each flake (rows 1–6): (a) reflected-light micrograph; (b) confocal telecom-band PL map (980-nm excitation; integration 1.50–1.60 μm); (c) AFM topography; (d) height line-profiles taken along the dashed lines in (c) (for Flake 2 and Flake 3, two orthogonal cuts "a" and "b" are shown). PL generally increases with thickness; bright ridges at some edges reflect geometry-dependent out-coupling.

*II.c Optical spectroscopy and lifetime measurements*

Optical spectra were recorded with a Princeton Instruments SpectraPro 2300i equipped with an InGaAs array and a 600 g/mm, 1600-nm blaze grating optimized for the 1.5-µm band. We first locate the emitting region by confocal imaging, then hold the 980-nm focus on the selected spot, collimate the collected light, and feed it to the spectrometer. Unless noted, each spectrum was integrated for 30 min. Time-resolved photoluminescence (PL) measurements were performed using a pulsed 980-nm diode laser driven by an iC-Haus Eval HB controller and triggered via a PulseBlaster (SpinCore). The sample was excited with a 200-µs laser pulse, after which the PL was recorded using a gated superconducting nanowire single-photon detector (SNSPD) with 50-µs time binning. The fluorescence decay profile was reconstructed by sweeping the delay time over a 10-ms interval.

**III Extended data sets**

*III.a Field-dependent Er PL (extends Main Fig. 2)*

Main Fig. 2 established that an applied magnetic field reduces the $Er^{3+}$ telecom PL and increases the excited-state lifetime, enabling us to use Er as a local magnetic probe. The results in Supplementary Fig. 4 expand on that result. We image Flake 3 while toggling a static field between $B = 0$ and 0.3 T and form ratio maps RPL = PL(0.3T)/PL(0T). The response is highly directional: An out-of-plane field ($B \parallel c$, the hardest magnetic axis) produces no discernible change, whereas in-plane fields dim the PL. Moreover, the dimming is largest for $B \parallel a$ and weaker for the orthogonal in-plane orientation. Time-resolved traces (not shown for brevity) confirm the anti-correlation introduced in the main text, namely, where PL decreases most, the lifetime $\tau$ grows most. The latter implies a reduction of the effective radiative rate rather than the creation of a fast non-radiative channel. Further, polarization-resolved data (see next sub-section below) show that under field the emission and excitation axes do not rotate; only the amplitude changes.

The spatial pattern of the dimming is typically non-uniform. While interiors show a relatively gentle, homogeneous reduction, edges and thickness steps frequently display enhanced suppression in the RPL maps. This edge sensitivity suggests that part of the transduction is photonic: The field alters CrSBr's magneto-optical response (birefringence/dichroism), which — in concert with local geometry — modifies the local density of optical states (LDOS) and the fraction of Er emission that couples into radiative vs guided/leaky modes. Edges and thickness gradients are where the LDOS is most field-susceptible, explaining the stronger local effect.

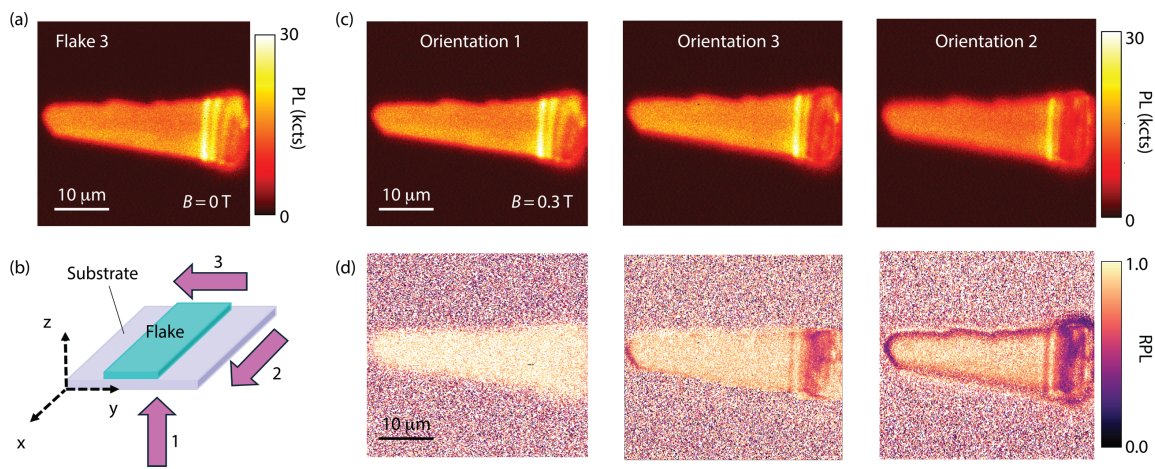

**Supplementary Figure 4: In- and out-of-plane field orientation dependence of $Er^{3+}$ PL in CrSBr.** (a) Confocal telecom-band PL map of Flake 3 at $B = 0$ T (980-nm excitation). (b) Schematic of the three field orientations used (arrows 1–3) with respect to the flake/substrate axes. (c) PL maps of Flake 3 at $B = 0.3$ T for orientations 1, 3, and 2 (left to right). The PL response is strongly directional: an out-of-plane field (B∥c) produces negligible change in brightness, whereas in-plane fields reduce the PL. The dimming is largest when B is aligned with the *a*-axis (intermediate magnetic axis) and weaker for the orthogonal in-plane orientation. (d) Ratio maps RPL = PL(0.3 T)/PL(0 T) visualize this anisotropy and also reveal that the suppression is often edge-enhanced, while the interior shows a smaller, more uniform reduction. All measurements at room temperature.

This said, a purely photonic explanation cannot account for the temperature-dependent PL/lifetime trends: To reproduce the dip near $T_N$ and the recovery deep in the AFM phase, the Er center must sense the local magnetic field itself, i.e., an "atomic" effect is mandatory. In this regime the relevant interaction volume is nanoscopic (set by the Er crystal-field wavefunction and exchange fields over ∼1–2 nm), so the readout reflects how well the two AFM sublattices cancel at the Er site and how much uncompensated moment resides in its immediate neighborhood.

At room temperature — well above $T_N$ — we cannot yet pin down why the dimming is strongest for $B \parallel a$. A conservative reading is that the anisotropy of the system sets the directionality, which could arise from any combination of anisotropic paramagnetic susceptibility, anisotropic magneto-optical response, and anisotropy of the Er crystal field/g-tensor — all different faces of the same underlying directional preference. Separately, given the evidence in Main Fig. 3 for residual AFM correlations well above $T_N$, it is also plausible that a field along $a$ cants these short-range moments more effectively, enhancing the local net field at Er; this mechanism is well-motivated at low $T$ and could persist at room temperature to the extent that such correlations survive.

### III.b Excitation and emission polarization

We first map the excitation dipole by rotating the pump polarization at $B = 0.0$ T, $T = 300$ K

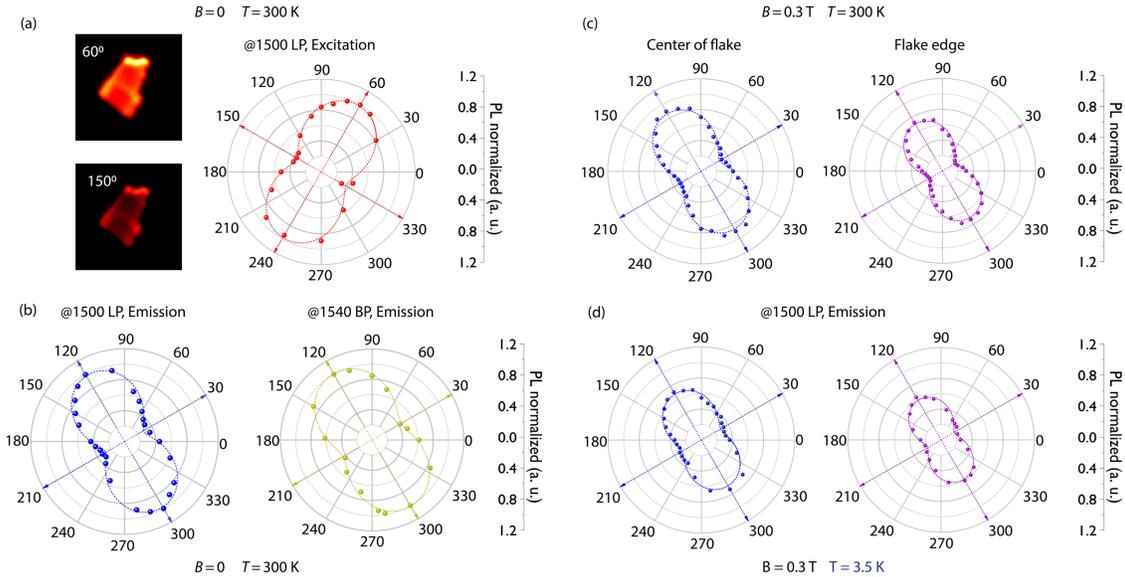

**Supplementary Figure 5: Polarization behavior of Er³⁺ emission in CrSBr.** (a) Excitation-angle series at $B = 0$, $T = 300$ K: confocal PL maps (left) and corresponding polar plots (right, 1500-nm long-pass) while rotating the pump linear polarization. PL is maximized when the pump is aligned with the flake's in-plane $a$-axis. (b) Emission polarization with excitation fixed at the optimal angle from (a): Integrated PL versus analyzer angle for a 1500-nm long-pass (left) and a 1540-nm band-pass (right), revealing strongly linearly polarized emission. (c) Effect of an in-plane field ($B = 0.3$ T) at $T = 300$ K: Analyzer-angle dependence recorded separately from the flake center (left) and edge (right) using a 1500-nm long-pass. The field reduces the overall amplitude but leaves the polarization axis unchanged; edge regions show a larger modulation. (d) Same measurement as (c) at $T = 3.5$ K. Curves are normalized to the maximum PL at 300 K and $B = 0$ T. Across all conditions we observe amplitude changes without rotation of the excitation or emission dipoles. In (c) and (d), the field points along the $a$-axis.

(Supplementary Fig. 5a). The integrated telecom PL follows an offset–cosine with a clear maximum when the pump aligns with the flake's in-plane *a*-axis, establishing that Er absorption is strongly linearly polarized. Fixing the pump at this optimum, we then rotate the analyzer (Supplementary Fig. 5b). Both a 1500-nm long-pass and a 1540-nm band-pass yield high-contrast, two-lobed polar plots with a stable analyzer axis, i.e., linearly polarized emission with little wavelength dependence across the dominant Er lines.

Applying an in-plane field $B = 0.3$ T reduces the amplitude of the polar plots but does not rotate either the excitation or emission axes (Supplementary Figs. 5c, 5d). The effect is temperature-robust (seen at 300 K and 3.5 K) and position dependent: Edge regions show a larger field-induced suppression than the flake interior (consistent with the observations in Fig. 2 of the main text). Combined, these observations indicate that the field primarily changes the magnitude of the radiative channel (or its out-coupling) rather than reorienting the optical dipole. The stronger edge response points to a photonic transduction component— edges and thickness gradients are where the LDOS/out-coupling are most sensitive—while the axis stability constrains purely atomic mechanisms to those that modulate oscillator strength without dipole rotation. This polarization behavior is therefore consistent with the two-stage picture developed in the main text: a local magnetic perturbation at the Er site, whose readout is amplified by the local photonic environment, especially near edges.

### III.c PL dependence on flake thickness (extends Main Figs. 3b–c)

In the main text we showed that the PL–temperature curves depend strongly on thickness: The thin (~70 nm) flake (Fig. 3b) exhibits a dip near $T_N$ with modest recovery of PL deep in the AFM phase, whereas the thick (~600 nm) flake (Fig. 3c) shows a similar dip but a much stronger low-*T* recovery. Supplementary Fig. 6 demonstrates the same trend within one sample (Flake 7) containing adjacent regions of different thickness. AFM confirms a clear height difference between Region 1 (thicker) and Region 2 (thinner); the corresponding PL–*T* traces reproduce the behavior above: The thinner region recovers comparatively more weakly below $T_N$, while the thicker region exhibits a larger PL rebound; note that the thickness difference within Flake 7 is smaller than seen within the flakes in the main text, consistent with a milder contrast

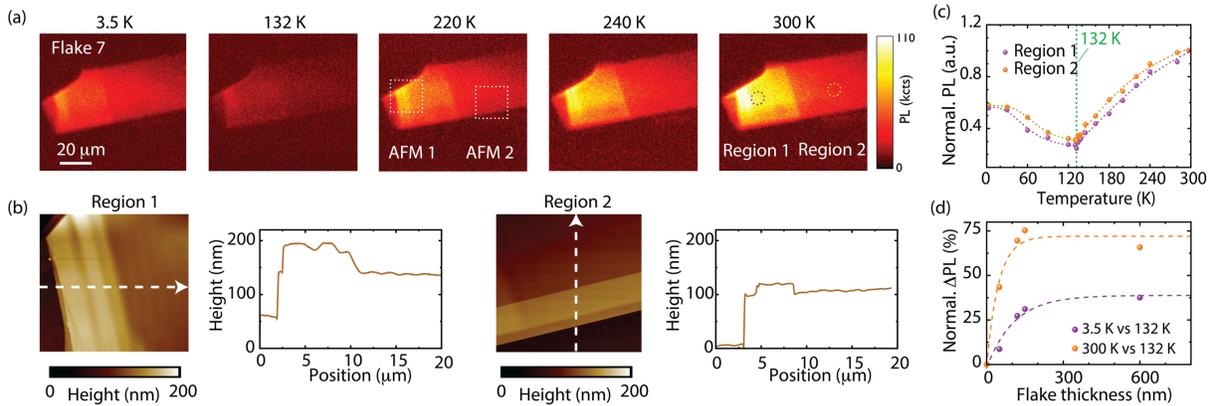

**Supplementary Figure 6: Thickness-dependent PL–temperature response within a single flake.** (a) Telecom-band PL maps of an Er-implanted CrSBr flake recorded while sweeping temperature (selected frames). Two areas of different thickness are indicated (Region 1: thicker; Region 2: thinner). (b) AFM topography and line profiles from the dashed boxes in (a) reveal the step in thickness between Region 1 and Region 2 (left/right panels, respectively). (c) Flake-integrated PL versus temperature extracted separately from the two regions under warm-up starting from 3.5 K. The thicker area (purple) reproduces the "thick-flake" behavior of the main text (deeper minimum and stronger recovery below $T_N$), whereas the thinner area (orange) follows the "thin-flake" trend (shallower minimum and more gradual evolution). The vertical dashed line marks the bulk Néel temperature $T_N \approx$ 132 K. (d) Normalized percentage difference $\Delta PL = \big(PL(T) - PL(132\,K)\big)/PL(300\,K)$ for temperatures $T = 3.5$ and 300 K (purple and orange circles, respectively) as derived from the data in (d) (and similar data sets for other flakes); dashed lines are guides to the eye. All data acquired with 980-nm excitation.

between curves.

Panel (d) extends this comparison across other flakes and quantifies this trend by plotting the normalized PL contrast $\Delta \text{PL} = \big(\text{PL}(T) - \text{PL}(132\text{ K})\big)/\text{PL}(300\text{ K})$ at $T = 3.5$ K and 300 K versus thickness; both low-$T$ and room-$T$ contrasts rise rapidly from ~70–150 nm and then saturate for thicker slabs. Combined, the intra-flake control (a–c) and cross-flake statistics (d) rule out flake-to-flake variability and show that the temperature response is systematically thickness-dependent. This finding seems to suggest that thicker slabs attain better sublattice cancellation (smaller net local field at the Er site).